\documentclass{article}
\usepackage{spconf,amsmath,graphicx}
\usepackage{cite}
\usepackage{amsmath,amssymb,amsfonts}
\usepackage{algorithmic}
\usepackage{graphicx}
\usepackage{textcomp}
\usepackage{soul}
\usepackage{algorithmic}
\usepackage[linesnumbered,ruled,vlined,noend]{algorithm2e}

\SetCommentSty{mycommfont}
\SetKwInput{KwInput}{Input}                
\SetKwInput{KwOutput}{Output}              
\SetKwInput{KwRequire}{Require}              

\usepackage[caption=true,font=footnotesize]{subfig}

\usepackage[dvipsnames]{xcolor}

\def\BibTeX{{\rm B\kern-.05em{\sc i\kern-.025em b}\kern-.08em
    T\kern-.1667em\lower.7ex\hbox{E}\kern-.125emX}}
\begin{document}
\title{Who is Snoring? Snore based User Recognition \\
}

\name{Shenghao Li,  Jagmohan Chauhan\vspace{-4.5mm}} 
\address{University of Southampton}
\maketitle
\begin{abstract} 
Snoring is one of the most prominent symptoms of Obstructive Sleep Apnea-Hypopnea Syndrome (OSAH), a highly prevalent disease that causes repetitive collapse and cessation of the upper airway.  Thus, accurate snore sound monitoring and analysis is crucial. However, the traditional monitoring method polysomnography (PSG) requires the patients to stay at a sleep clinic for the whole night and be connected to many pieces of equipment. An alternative and less invasive way is passive monitoring using a smartphone at home or in the clinical settings. But, there is a challenge: the environment may be shared by  people such that the raw audio may contain the snore activities of the bed partner or other person. False capturing of the snoring activity could lead to critical false alarms and misdiagnosis of the patients. To address this limitation, we propose a hypothesis that snore sound contains unique identity information which can be used for user recognition. We analyzed various machine learning models: Gaussian Mixture Model (GMM), GMM--UBM (Universal Background Model), and a Deep Neural Network (DNN) on MPSSC - an open source snoring dataset to evaluate the validity of our hypothesis. Our results are promising as we achieved  around or more than 90\% accuracy in  identification and verification tasks.  This work marks the first step towards understanding the practicality of snore based user monitoring to enable  multiple healthcare applications.  


\end{abstract}


\section{Introduction}  
Snoring is a common symptom of sleep disorder breathing (SDB). It is caused by repeated bouts of decreased and stopped airflow during sleep. According to an earlier investigation, about 50\% of the adult population frequently snores \cite{10.1093/sleep/3.3-4.221}. It has become a significant reason for poor sleep quality, degraded memory, and diseases such as sleep disturbance and high blood pressure \cite{8857884, zhang2019snoregans}. Previous studies have revealed that analysis of snore sound is crucial for respiratory and cardiopulmonary diseases at an early stage. Detected snores can be further analyzed to extract more detailed attributes such as upper airway condition. Thus, snore sound analysis is  vital  for  passive health monitoring systems.

The most prevalent and reliable technique for snoring sound monitoring and diagnosis is polysomnography (PSG) \cite{9053683}. Yet it is also confounded by problems like long reservation lists, high nursing costs, and the "first-night effect" due to environmental changes \cite{8857884}. Recently, acoustic monitoring using a smartphone at home or ward has aroused considerable interest for its non-invasiveness and  cost-effectiveness. However, this raise an issue if the space is shared by a bed partner or other patients since the raw audio may contain interfering snore sounds which can cause false alarms and impact the  performance of the health monitors. 

Previous studies have focused on snore sound classification of the four types of snoring sounds (Velum, Oropharyngeal, Tongue, and Epiglottis) based on their excitation source. For example,  Wang et  al. \cite{8553521} proposed an architecture based on Convolutional Neural Network combined with Gated Recurrent Unit (CNN-GRU) to extract features solely from the training set in the corpus, and then processed them through the Channel Slice Model to obtain an Unweighted Average Recall (UAR) of 63.8\%. Kun Qian et al. \cite{Qian2016ClassificationOT} carried out a multi-feature analysis of the snore sounds. They applied those features to several classifiers to rank the features based on their importance and achieved an UAR of 78\%.

In order to analyze the snoring sound in a shared environment, we must deal with the snorer recognition problem in the first place. Inspired by the idea that the feature of a subject's cough is determined by the lung tissue and vocal cord \cite{Larson:2011:APP:2030112.2030163}, we assume that the snoring sound may contain a unique signature, which can be utilized to determine who actually snored. In this paper, our main aim is to determine if users can be identified/verified on the basis of their snore sounds. To this end, we  created machine learning models based on  GMM, GMM-UBM, and Deep Neural Network (DNN) to extract snore patterns and evaluated them on MPSCC data corpus.  Our results  are promising  as   all three classifiers achieved an identification rate of more than 90\% and less than 15\% equal error rate (EER) with just four snoring during the enrollment per user. As for verification performance, snore embeddings derived from DNN produce the best results by achieving over 95\% specificity and sensitivity.  Our work  is the first to  show that snoring sounds  have unique signatures and can  potentially be used for user monitoring in shared environments. 




\section{Related Work}
Existing studies have focused on  using audio such as speech and cough for user authentication. However, no previous study  has investigated if snore has unique signature which can determine who produced it. As for speaker recognition,  Automatic Speaker Recognition (ASR) system aims at eliciting, classifying, and recognizing the voice signal that conveys speaker identity \cite{5745552}. Modern approaches are mainly based on the statistical models of speech signal. Reynolds et al. \cite{365379} proposed an individual speaker identification method based on GMM.  Later in 2000, Reynolds \cite{Reynolds2000SpeakerVU} proposed a modified GMM with a Universal Background Model (UBM) for the adaption of each enrolled subject .  To decrease the complexity of the GMM supervector, Dehak et al. \cite{5545402} exploited  i-vectors to train the classifiers and identify unknown speakers. Deep Learning methods such as x-vectors have also been proposed  for voice print recognition \cite{8461375}. The embeddings derived from x-vectors outperformed the i-vectors. 

For cough recognition, Whitehill et al. \cite{9053268} implemented a multitask learning system and achieved an identification accuracy of 82.15\% . The model was trained on both speech and cough to address the lack of sufficient training data.  Vatanparvar et al. \cite{9176835} proposed a novel cough embedding model to do cough-based user verification. A four-layer dense Neural Network is applied to the training set of cough signals. The model achieved an average error rate of below 10\% with only five utterances per subject in the enrollment phase. Recently, Cleres et al. \cite{jokic2022tripletcough} applied a triplet network architecture based on CNN and proposed a metric learning algorithm for cough recognition. With the audio recordings of asthmatic patients’ coughs, their method produced an accuracy  of 80\% and an EER of 20\% using 12 utterances in the enrollment phase.

\section{Methodology}
\subsection{Data Set, PreProcessing and Feature Extraction}

We used MPSSC (Munich-Passau Snore Sound Corpus) dataset \cite{schuller2017interspeech} for our work.  This corpus  obtained primary material from Drug-Induced Sleep Endoscopy (DISE) examinations. After manual selection, the dataset only contains events without interference and with a discernible, single point of vibration. Snore sounds are divided into four categories based on their excitation location in the upper airways. 828 utterances were collected from 219 subjects.  The audio was recorded with a sampling rate of 16KHz with 16-bit linear PCM mono channel.

\begin{figure}[t]
\centering
\subfloat[Subject 1]{
\includegraphics[width=4.75cm]{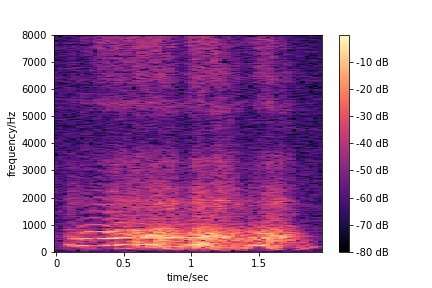}
}
\subfloat[Subject 2]{
\includegraphics[width=4.75cm]{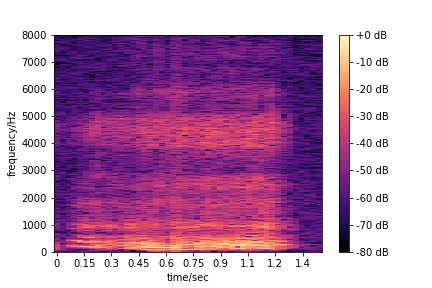}
}
\caption{Spectrogram of two snorers}
 \label{fig:stft}
\end{figure}

 We applied  Short-time Fourier transform (STFT) on the audio signals to represent the signal in the time-frequency domain by computing discrete Fourier transform (DFT) over short overlapping windows. The audio was  divided into 25-ms frames by a Hanning window, with 15 ms of overlap. 25 MFCC coefficients  were extracted   as  feature inputs for all the three models. Figure \ref{fig:stft} shows the spectrogram of two  snorers.

\subsection{Model Design}
Here we explain three models (GMM, GMM-UBM, DNN) for snore recognition along with their training and evaluation. 
\\
\textbf{GMM}: Gaussian Mixture Model (GMM) is efficient in signal approximation because it can fit complex probability distribution of time series by a small number of parameters performing a weighted average of multiple simple normal distributions. According to Reynold et al.\cite{365379}, GMM offers a reliable speaker representation for the challenging task of speaker recognition using distorted, unrestricted speech and enables high recognition accuracy for short utterances. So, we tried GMM with different components in our work. 



Since the MPSSC corpus has limited utterances (snoring samples) for each person, we select the subjects with at least five utterances/samples. 72 subjects meet our requirements. The first four samples are used for training and the other one for testing.   In the training phase, we first extract the frame-level feature vectors and stack the ones of the same subject together to form an aggregated feature matrix.   We model all the snorers independently from the extracted features. The model for each subject is constructed by fitting the distribution with the feature matrix. For the evaluation phase, for each sample, the model with the highest likelihood score is considered as the identified snorer (identification) or the likelihoods are compared to a threshold to determine if the sample matches the model for verification.
\\
\textbf{GMM-UBM}: It is an extension of the Gaussian Mixture model. It is particularly effective for users with limited enrollment speech samples. The core is to employ other users' data for pre-training and reduce the dependence on the subject data. We first need to train a background Gaussian Mixture Model based on all the user's data, then fit each Gaussian distribution to the target user's data. This procedure is called adaptation using Maximum a posterior (MAP) algorithm \cite{Reynolds2000SpeakerVU}. We can fuse the obtained new parameters with the original parameters of the UBM model to get the target speaker's model.  We use the same concepts in our work for the snore sounds. Similar to the evaluation phase for the GMM model, we also calculated the log-likelihood of a given utterance from each model and choose the one with the highest score as the predicted subject. For the verification, the decision is made through a preset threshold. 

\textbf{DNN}: Inspired by the success of Deep Neural Network (DNN) on  speaker verification and cough recognition  \cite{6854363,9176835}, we used DNN as a snore embedding extractor. We used the same methodlogy as proposed in CoughMatch \cite{6854363} by dividing the task  into three parts. In the development phase, we train a supervised DNN, based on the frame level, with all the subjects in the dataset. For the subject that has more than four utterances, we use the first four. Otherwise, we use all its audio.
 we assume the training utterance from the snorer s as:
\begin{equation}
    X_{s}=\left \{ O_{s1},O_{s2},\dots ,O_{sn} \right \}
\end{equation}
Each utterance is represented by several selected observations (frames):
\begin{equation}
    O_{si}=\left \{ o_{1},o_{2},\cdots ,o_{m} \right \}
\end{equation}
The network's input is formed by stacking each observation's 25-dimensional MFCC features vector with the ones of its 49 adjacent frames. Thus, the dimension of the input vector is 1250.  The output is a 1-hot N-dimensional identity vector, where the only non-zero item is the one corresponding to the subject identity. N is the number of the subject. 
 
 In the enrollment phase, 72 subjects with 5 utterances are used for the embedding extraction. We freeze all the dense layers and remove the output layer. The input feature is generated in the same approach as described above. For each subject, an independent snore embedding is constructed with four training utterances.  Here, each utterance contains 15 selected observations (frames), the distance between adjacent observations is five frames:
\begin{equation}
    O_{ei}=\left \{ o_{1},o_{2},\cdots ,o_{15} \right \}
\end{equation}
We feed every observation with its 50 neighboring frames to the supervised trained DNN; L2-normalize the output obtained from the last hidden layer, and accumulate for all the observations from utterance $O_{si}$. The outcome vector is referred to as the utterance-level snore embedding. Finally, the subject's snore embedding is acquired by averaging the embeddings of the four utterances in $X_{s}$.

In the evaluation phase, another utterance from each of the 72 subjects is used for testing. The testing sample is strictly excluded from either the development set or the enrollment set. We implement the same feature extraction (feeding it to the DNN) for the testing utterance and compare the cosine similarity between the test embedding and the claimed subject's embedding. The decision is made on the subject that has the highest similarity (identification) or comparing the similarity to a threshold (verification). We created a 4-layer network with 128 nodes per layer and adopt rectified linear units (Relu) as the activation function for hidden layers. We adopt 'Softmax' as the activation function for the output layer and insert a 15\% Dropout before the output layer to mitigate the overfitting problem. The network is trained for 80 epochs with Adam optimizer.

\section{Results}
\subsection{Snore Signature}
\begin{figure}[t]
\centering
\subfloat[MFCC features]
{
\includegraphics[width=4.75cm]{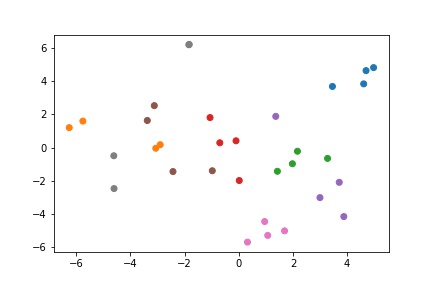}
}
\subfloat[Snore embeddings from DNN]{
\includegraphics[width=4.75cm]{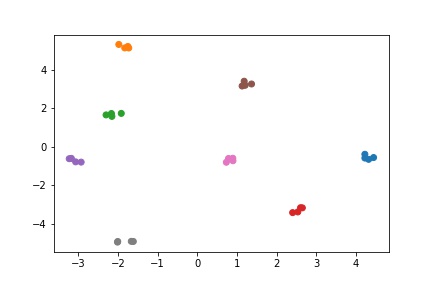}
}
 
\caption{T-SNE Plots}
 \label{fig:tsne}
\end{figure}

We firstly show the visual patterns of snores obtained from different users.  Fig. \ref{fig:tsne}  shows the t-sne plots \cite{JMLR:v9:vandermaaten08a}. The left hand side shows the distribution of MFCC features of the first eight subjects. Each point represents an utterance from the training set, and the point's color represents a certain subject. The distribution of the same eight subjects' snore embeddings obtained from DNN is shown on the right side. We only show eight to ensure the readability of the plot.     It can be seen that the snore embeddings have better clustering performance over the MFCC features. This also shows that snore have unique signature which varies between users and can be useful for snore tracking and user recognition.

\begin{table}[!htb]
  \centering
  \caption{GMM and GMM-UBM identification accuracy}
  \begin{tabular}{|c|c|c|} \hline
Number  of Components & GMM & GMM-UBM \\ \hline
  5 & 0.9722 &   1.0000 \\ \hline
  10 & 0.9722  & 0.9861 \\ \hline
  15 & 0.9583   & 0.9722 \\ \hline
  20 & 0.9583   & 0.9722 \\ \hline
  25 & 0.9444   & 0.9583 \\ \hline
  \end{tabular}
  \label{table1}
\end{table}



\begin{figure*}[t]
\centering
\subfloat[GMM]
{
\includegraphics[width=0.3\textwidth]{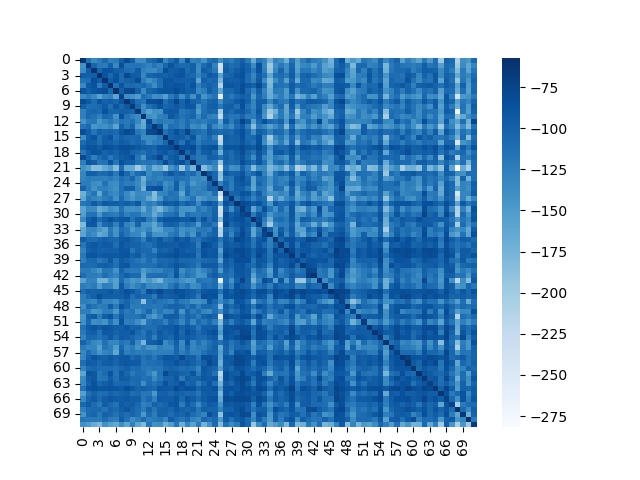}
}
\quad
\subfloat[GMM-UBM]{
\includegraphics[width=0.3\textwidth]{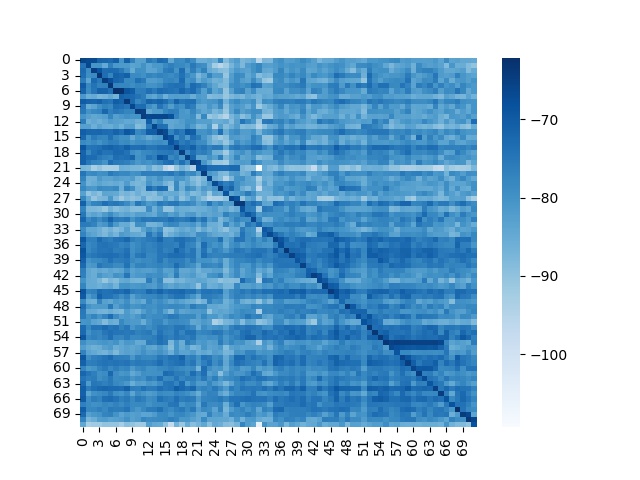}
}
\quad
\subfloat[DNN embeddings]{
\includegraphics[width=0.3\textwidth]{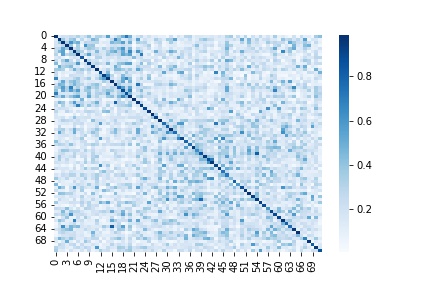}
}
\caption{ Heatmap of the three models - Verification }
 \label{fig:heatmap}
\end{figure*}

\subsection{Results for Identification}
We obtain more than 90\% accuracy with the GMM based models as shown in Table \ref{table1}. Theoretically, the fitting effect should be promoted with the increment of Gaussian components.  However as the number of training samples were few, the  accuracy  decreases with the number of  components.  


\begin{figure}[t]
\centering
\subfloat[GMM]
{
\includegraphics[width=4.25cm]{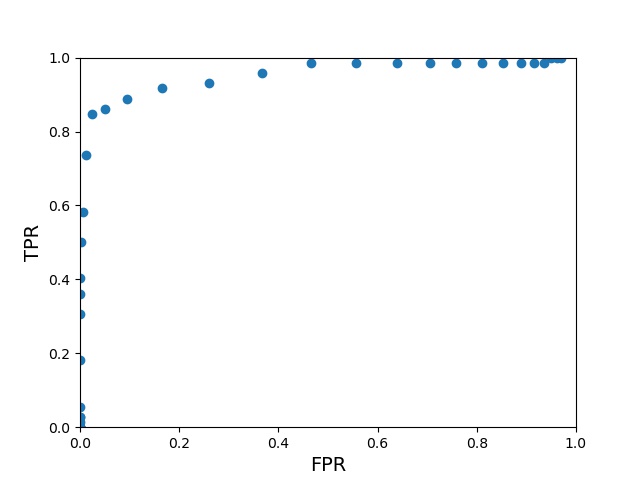}
}
\subfloat[GMM-UBM]{
\includegraphics[width=4.25cm]{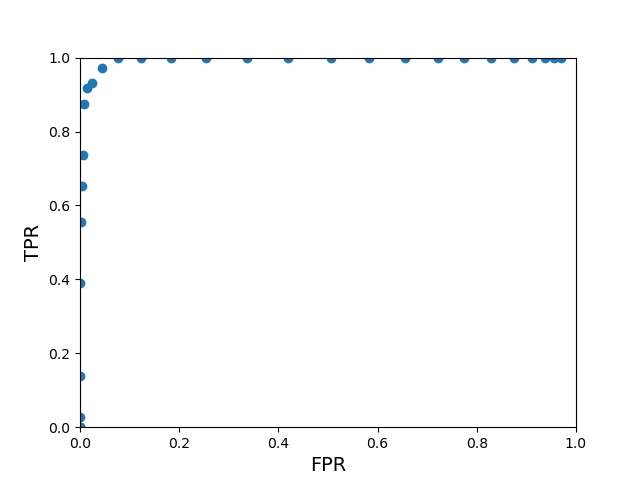}
}
\quad
\subfloat[DNN embeddings]{
\includegraphics[width=4.25cm]{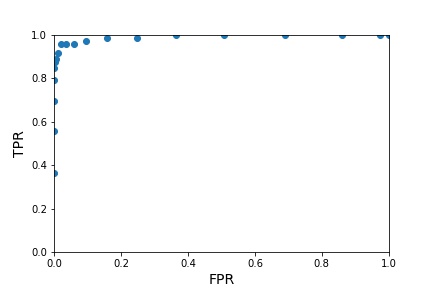}
}
\caption{ROC of the three models - Verification}
 \label{fig:roc}
\end{figure}

GMM-UBM model improves the system performance by pre-training a universal background model. The identification accuracy over the number of components  is shown in Table \ref{table1}. The accuracy of GMMM-UBM at each component  gets slightly better than a GMM model which verifies the superiority of GMM-UBM over GMM. In fact, the accuracy ranges from 95\% to 100\% for GMM-UBM.   For the DNN, we obtained an  accuracy of 0.9306 with snore embeddings.
\subsection{Results for Verification}
For user verification, we set the number of the Gaussian components at 10 as that obtained the best performance in our analysis.
We initially visualize the verification performance intuitively, using the heatmap, for the three  models successively (see Figure. \ref{fig:heatmap}). The row index denotes the test number, while the column index denotes the embedding number. Item color varies with the evaluation value. Thus the ideal case is that the item on the diagonal line has the deepest shade if all the test audios are correctly verified. It can be seen visually that the snore embedding extracted from the DNN has the best verification performance.

We analyzed the verification performance using different metrics. Firstly, we  compare  the specificity (True Negative Rate) and sensitivity (True Positive Rate) values satisfying an equal error rate (EER). We adjust the threshold for each model to get the best performance (see Table \ref{table2}). As shown, the snore embeddings from the DNN model achieve the best results over the other two models based on GMM.  Next, we calculated the receiver operating characteristic curve (ROC) to evaluate the verification performance. The bigger the area under the curve, the better the result is. ROC curves of the three models is illustrated in Figure \ref{fig:roc}. Similar to our previous result, the DNN based model has the highest TPR rate (sensitivity) and the smallest TPR-TNR distance. Thus, it achieves the best verification result among all the three models.

\begin{table}[!htb]
  \centering
  \caption{Verification performance for the three models  }
  \begin{tabular}{|c|c|c|c|c|} \hline
  Model & Threshold & TPR & TNR  \\ \hline
  GMM & -33.2 & 0.9028 & 0.8940\\ \hline
  GMM-UBM & -71.1 & 0.9444 & 0.9701\\ \hline
  DNN & 0.45 & 0.9583 & 0.9523\\ \hline
  \end{tabular}
  \label{table2}
\end{table}



\section{Discussion and Conclusions} 
Passive health monitoring  of snore sounds using ubiquitous devices is becoming prevalent due to its non-invasiveness and low cost. However, such monitoring will fail in   scenarios where  many people share the same environment such as in a house. Thus, snorer recognition is a subject worthy of study. This paper evaluated three models for snorer identification and verification based on the MPSSC corpus. MFCC features are extracted to create each subject's snore model, which is later used to check if the testing utterance can match the individual snore pattern. We  mitigated the overfitting problem due to  limited training data by introducing a Universal Background Model. A Deep Neural Network is also applied to train the snore embeddings to improve the system's performance. Our  results are encouraging and show that the models can achieve over 90\% identification accuracy.  As for verification results, the snore embedding model based on DNN obtains over 95\%  for sensitivity and specificity. To conclude,  our work takes  the first ever step towards understanding the practicality  and proving the  feasibility of user recognition through snores that can lead to  creation of  accurate and ubiquitous  snore monitoring systems in the future. 
\bibliographystyle{IEEEtran}

\bibliography{mybib}

\begin{thebibliography}{10}
\providecommand{\url}[1]{#1}
\csname url@samestyle\endcsname
\providecommand{\newblock}{\relax}
\providecommand{\bibinfo}[2]{#2}
\providecommand{\BIBentrySTDinterwordspacing}{\spaceskip=0pt\relax}
\providecommand{\BIBentryALTinterwordstretchfactor}{4}
\providecommand{\BIBentryALTinterwordspacing}{\spaceskip=\fontdimen2\font plus
\BIBentryALTinterwordstretchfactor\fontdimen3\font minus
  \fontdimen4\font\relax}
\providecommand{\BIBforeignlanguage}[2]{{%
\expandafter\ifx\csname l@#1\endcsname\relax
\typeout{** WARNING: IEEEtran.bst: No hyphenation pattern has been}%
\typeout{** loaded for the language `#1'. Using the pattern for}%
\typeout{** the default language instead.}%
\else
\language=\csname l@#1\endcsname
\fi
#2}}
\providecommand{\BIBdecl}{\relax}
\BIBdecl

\bibitem{10.1093/sleep/3.3-4.221}
\BIBentryALTinterwordspacing
E.~Lugaresi, F.~Cirignotta, G.~Coccagna, and C.~Piana, ``{Some Epidemiological
  Data on Snoring and Cardiocirculatory Disturbances},'' \emph{Sleep}, vol.~3,
  no. 3-4, pp. 221--224, 09 1980. [Online]. Available:
  \url{https://doi.org/10.1093/sleep/3.3-4.221}
\BIBentrySTDinterwordspacing

\bibitem{8857884}
J.~Sun, X.~Hu, Y.~Zhao, S.~Sun, C.~Chen, and S.~Peng, ``Snorenet: Detecting
  snore events from raw sound recordings,'' in \emph{2019 41st Annual
  International Conference of the IEEE Engineering in Medicine and Biology
  Society (EMBC)}, 2019, pp. 4977--4981.

\bibitem{zhang2019snoregans}
Z.~Zhang, J.~Han, K.~Qian, C.~Janott, Y.~Guo, and B.~Schuller, ``Snore-gans:
  Improving automatic snore sound classification with synthesized data,'' 2019.

\bibitem{9053683}
H.~E. Romero, N.~Ma, and G.~J. Brown, ``Snorer diarisation based on deep neural
  network embeddings,'' in \emph{ICASSP 2020 - 2020 IEEE International
  Conference on Acoustics, Speech and Signal Processing (ICASSP)}, 2020, pp.
  876--880.

\bibitem{8553521}
J.~Wang, H.~Strömfeli, and B.~W. Schuller, ``A cnn-gru approach to capture
  time-frequency pattern interdependence for snore sound classification,'' in
  \emph{2018 26th European Signal Processing Conference (EUSIPCO)}, 2018, pp.
  997--1001.

\bibitem{Qian2016ClassificationOT}
K.~Qian, C.~Janott, V.~Pandit, Z.~Zhang, C.~Heiser, W.~Hohenhorst, M.~Herzog,
  W.~Hemmert, and B.~Schuller, ``Classification of the excitation location of
  snore sounds in the upper airway by acoustic multi-feature analysis.''
  \emph{IEEE transactions on bio-medical engineering}, 2016.

\bibitem{Larson:2011:APP:2030112.2030163}
\BIBentryALTinterwordspacing
E.~C. Larson, T.~Lee, S.~Liu, M.~Rosenfeld, and S.~N. Patel, ``Accurate and
  privacy preserving cough sensing using a low-cost microphone,'' in
  \emph{Proceedings of the 13th International Conference on Ubiquitous
  Computing}, ser. UbiComp '11.\hskip 1em plus 0.5em minus 0.4em\relax New
  York, NY, USA: ACM, 2011, pp. 375--384. [Online]. Available:
  \url{http://doi.acm.org/10.1145/2030112.2030163}
\BIBentrySTDinterwordspacing

\bibitem{5745552}
D.~A. Reynolds, ``An overview of automatic speaker recognition technology,'' in
  \emph{2002 IEEE International Conference on Acoustics, Speech, and Signal
  Processing}, vol.~4, 2002, pp. IV--4072--IV--4075.

\bibitem{365379}
D.~Reynolds and R.~Rose, ``Robust text-independent speaker identification using
  gaussian mixture speaker models,'' \emph{IEEE Transactions on Speech and
  Audio Processing}, vol.~3, no.~1, pp. 72--83, 1995.

\bibitem{Reynolds2000SpeakerVU}
D.~A. Reynolds, T.~F. Quatieri, and R.~B. Dunn, ``Speaker verification using
  adapted gaussian mixture models,'' \emph{Digit. Signal Process.}, vol.~10,
  pp. 19--41, 2000.

\bibitem{5545402}
N.~Dehak, P.~J. Kenny, R.~Dehak, P.~Dumouchel, and P.~Ouellet, ``Front-end
  factor analysis for speaker verification,'' \emph{IEEE Transactions on Audio,
  Speech, and Language Processing}, vol.~19, no.~4, pp. 788--798, 2011.

\bibitem{8461375}
D.~Snyder, D.~Garcia-Romero, G.~Sell, D.~Povey, and S.~Khudanpur, ``X-vectors:
  Robust dnn embeddings for speaker recognition,'' in \emph{2018 IEEE
  International Conference on Acoustics, Speech and Signal Processing
  (ICASSP)}, 2018, pp. 5329--5333.

\bibitem{9053268}
M.~Whitehill, J.~Garrison, and S.~Patel, ``Whosecough: In-the-wild cougher
  verification using multitask learning,'' in \emph{ICASSP 2020 - 2020 IEEE
  International Conference on Acoustics, Speech and Signal Processing
  (ICASSP)}, 2020, pp. 896--900.

\bibitem{9176835}
K.~Vatanparvar, E.~Nemati, V.~Nathan, M.~M. Rahman, and J.~Kuang, ``Coughmatch
  – subject verification using cough for personal passive health
  monitoring,'' in \emph{2020 42nd Annual International Conference of the IEEE
  Engineering in Medicine \& Biology Society (EMBC)}, 2020, pp. 5689--5695.

\bibitem{jokic2022tripletcough}
S.~Joki{\'c}, D.~Cleres, F.~Rassouli, C.~Steurer-Stey, M.~A. Puhan,
  M.~Brutsche, E.~Fleisch, and F.~Barata, ``Tripletcough: Cougher
  identification and verification from contact-free smartphone-based audio
  recordings using metric learning,'' \emph{IEEE Journal of Biomedical and
  Health Informatics}, vol.~26, no.~6, pp. 2746--2757, 2022.

\bibitem{schuller2017interspeech}
B.~Schuller, S.~Steidl, A.~Batliner, E.~Bergelson, J.~Krajewski, C.~Janott,
  A.~Amatuni, M.~Casillas, A.~Seidl, M.~Soderstrom \emph{et~al.}, ``The
  interspeech 2017 computational paralinguistics challenge: Addressee, cold \&
  snoring,'' in \emph{Computational Paralinguistics Challenge (ComParE),
  Interspeech 2017}, 2017, pp. 3442--3446.

\bibitem{6854363}
E.~Variani, X.~Lei, E.~McDermott, I.~L. Moreno, and J.~Gonzalez-Dominguez,
  ``Deep neural networks for small footprint text-dependent speaker
  verification,'' in \emph{2014 IEEE International Conference on Acoustics,
  Speech and Signal Processing (ICASSP)}, 2014, pp. 4052--4056.

\bibitem{JMLR:v9:vandermaaten08a}
\BIBentryALTinterwordspacing
L.~van~der Maaten and G.~Hinton, ``Visualizing data using t-sne,''
  \emph{Journal of Machine Learning Research}, vol.~9, no.~86, pp. 2579--2605,
  2008. [Online]. Available:
  \url{http://jmlr.org/papers/v9/vandermaaten08a.html}
\BIBentrySTDinterwordspacing

\end{thebibliography}
\end{document}